# Signatures of Technetium Oxidation States: A New Approach

Stephen Bauters[a,b], Andreas C. Scheinost[a,b], Katja Schmeide[b], Stephan Weiss[b], Kathy Dardenne[c], Jörg Rothe[c], Natalia Mayordomo[b], Robin Steudtner[b], Thorsten Stumpf[b], Ulrich Abram[d], Sergei M. Butorin[e,*] and Kristina O. Kvashnina[a,b,*]

A first general strategy for the determination of Tc oxidation state by new approach involving X-ray absorption near edge spectroscopy (XANES) at the Tc $L_3$ edge is shown. A comprehensive series of $^{99}$Tc compounds, ranging from oxidation states I to VII, was measured and subsequently simulated within the framework of crystal-field multiplet theory. The observable trends in absorption edge energy shift in combination with the spectral shape allow for a deeper understanding of complicated Tc coordination chemistry. This approach can be extended to numerous studies of Tc systems as this method is one of the most sensitive methods for accurate Tc oxidation state and ligand characterization.

Technetium is the chemical element with atomic number 43, situated between manganese and rhenium in group 7 of the periodic table. As predicted by the periodic law, its properties are in between those two elements, but its chemistry is more similar to rhenium. Discovered in 1937, it is the lightest element for which all isotopes are radioactive.[1]. $^{99}$Tc is primarily produced in bulk quantities in nuclear reactors and is considered as unwanted radioactive waste. The short-lived metastable nuclide $^{99m}$Tc ($\gamma$-emitter, $t_{1/2}$ = 6.01 h) is the dominating isotope in routine diagnostic nuclear medicine with approximately 40 million administrations annually.

Technetium exhibits nine oxidation states from −I to VII, with IV, V, and VII being the most common.[2,3] Lower Tc oxidation states are commonly supported by $\pi$-acceptor ligands such as isocyanides, CO or $NO^+$. It has been shown that $Tc^V$, $Tc^{III}$ and $Tc^I$ compounds exhibit a diverse chemistry and are frequently used in nuclear medicine.[4–7] Investigating these types of compounds and their specific coordination chemistry is crucial to determine and improve both their stability and their biological activity.[4] Technetium compounds in the II and VI oxidation states are relatively rare in literature.[2,3,8–10] In general, studies to identify Tc oxidation states using other techniques often do not probe the Tc oxidation state directly, but rather determine oxidation states based on structure parameters.[2,3,11]

A well-defined oxidation state in coordination species is of paramount importance, since ligand atoms can be connected differently to the coordination center, depending on its oxidation state. The assignment of the oxidation state is generally based on the crystal structure and stoichiometry, on spectroscopic analysis, on magnetic properties or on computational modelling results. The practical difficulty is, however, to establish a methodology for the direct Tc oxidation state characterization, which was hitherto considered a challenging task.[12] Spectroscopic techniques such as Tc K edge XANES or Raman spectroscopy (c.f. ESI) can provide complementary information about ligands and their coordination, but ultimately lack the sensitivity when compared to Tc $L_3$ edge XANES.

We report here the first direct and comprehensive measurements of the Tc oxidation state on a variety of Tc compounds, which were achieved by the XANES technique at the Tc $L_3$ edge (~2677 eV). The main goal of this paper is to show a systematic comparison of the Tc $L_3$ XANES data for an extensive series of Tc coordination species with oxidation states extending from I to VII as derived by their crystal structure and stoichiometry. To strengthen our conclusions and to make the assignments, we compare a series of experimental spectra with simulated ones by crystal-field multiplet theory.

$^{99}$Tc is a long-lived weak $\beta^-$ emitter ($E_{max}$ = 0.292 MeV). Normal glassware provides adequate protection against the weak beta radiation when milligram amounts are used. Secondary X-rays (bremsstrahlung) play a significant role only when larger amounts of $^{99}$Tc are handled. All manipulations were done in a laboratories licensed for handling of $^{99}$Tc.

All chemicals were reagent grade and used without further purification. $^{99}$Tc was purchased as solid ammonium pertechnetate from Oak Ridge National Laboratory (USA). The salt was purified by recrystallization from aqueous solution. $KTcO_4$ was prepared by dissolution of equivalent amounts of $NH_4TcO_4$ and KOH in hot water. The less soluble potassium salt precipitates upon cooling and can be recrystallized from water when required. The syntheses of $(NBu_4)[TcNCl_4]$,[13] $(NBu_4)[TcNBr_4]$,[13] $(NBu_4)[TcOCl_4]$,[14] $(NBu_4)[TcOBr_4]$,[15] $[TcCl_4(PPh_3)_2]$,[16] $[TcCl_3(PPh_3)_2(CH_3CN)]$,[17] $[TcCl_3(PMe_2Ph)_3]$,[18] $[Tc(thio-urea)_6]Cl_3$,[19] $(NBu_4)[Tc(NO)Br_4]$,[20] $(NBu_4)[Tc(NO)Cl_4(MeOH)]$,[21] $[Tc(NO)Cl_2(PPh_3)_2(CH_3CN)]$[22] and $(NBu_4)[Tc_2(CO)_6Cl_3]$[23] followed published procedures. For the syntheses of $(NH_4)_2[TcCl_6]$ and $(NH_4)_2[TcBr_6]$, solutions of $(NH_4)TcO_4$ in HCl (37%) or HBr (48%) were heated under reflux for

10 min. (NH$_4$)Cl or (NH$_4$)Br were added and the products precipitated upon cooling as yellow or red crystals.

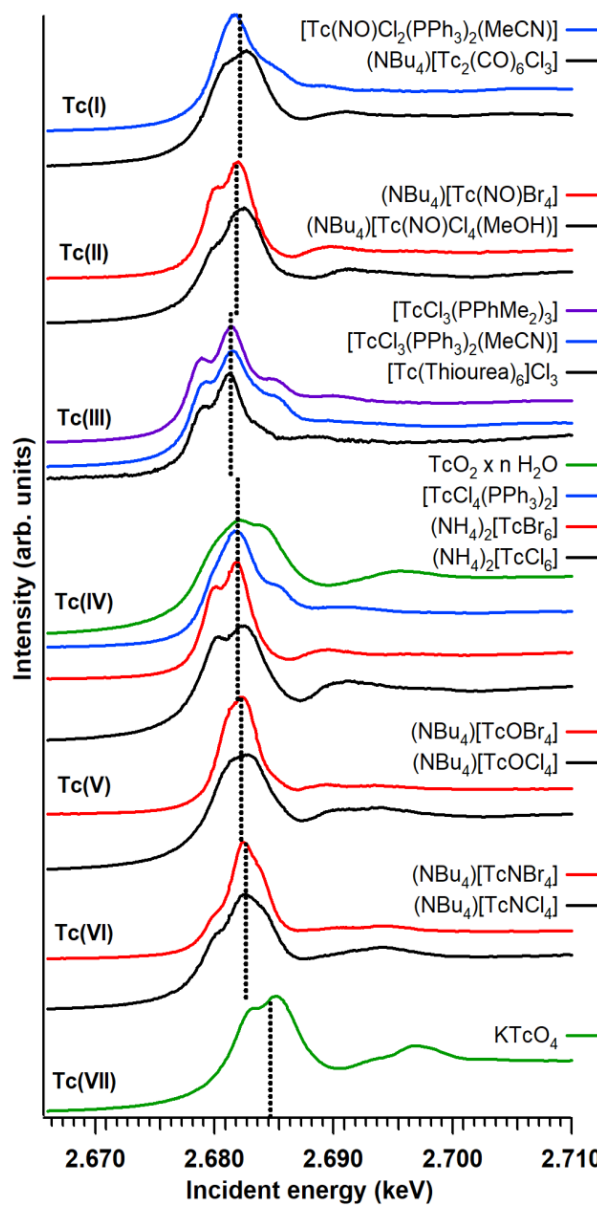

**Fig. 1**. Experimental Tc L$_3$ edge XANES data for the Tc compounds. The dotted lines mark the white line position as determined with the center of gravity

The Tc L$_3$ edge XANES spectra were recorded at the INE-Beamline of the KARA (Karlsruhe research accelerator) facility in Karlsruhe, Germany.[24] This bending magnet beamline operates with a collimating first mirror and a toroidal refocusing second mirror, both Rh-coated to reduce higher harmonics and to condense the photon flux at the sample position, and a fixed-exit double crystal monochromator.[25] An energy range from 2635 eV to 2823 eV was scanned using Si(111) crystals with a varying step size down to 0.1 eV in the edge region. The incoming flux was measured using an ionization chamber (Oken, Japan) with a windowless attachment to the He-filled sample chamber. The solid samples were sealed in sample holders, protected by 8 μm Kapton windows. Each sample holder was placed at a 45° angle to the incoming beam inside the sample chamber. We tried to minimize self-absorptions effects by collecting data at grazing exit angle (comparison is shown in Fig.S1). Since we cannot fully exclude remaining self-absorption, we do not consider the influence of oxidation state on the white line amplitude in the present investigation. A standard 90° fluorescence detection geometry (Fig.S2) was employed for the single element SDD (silicon drift detector) Vortex detector (Hitachi, USA). The SDD detector was separated from the He environment by a thin Kapton window to prevent He diffusion through the Be-window due to prolonged He exposure, which would degrade the internal detector vacuum and functioning.

The spectral calculations for the Tc systems were performed within the framework of crystal-field multiplet theory in a common manner as described before, e.g.[26,27] The TT-MULTIPLETS program package was used. Slater integrals, spin-orbit coupling constants and matrix elements for the Tc $4d^n \rightarrow 2p^5 4d^{n+1}$ transitions (where n was varied from 0 to 5) were obtained using Cowan's[28] and Butler's[29] codes, respectively, which have been modified by Thole[30].

| Compound | Approximated local symmetry | Crystal-field parameters (eV) |
|---|---|---|
| (NBu$_4$)[Tc(NO)Cl$_4$(MeOH)] | O$_h$ | 10Dq = 3.5 |
| [Tc(thiourea)$_6$]Cl$_3$ | O$_h$ | 10Dq = 3.5 |
| (NH$_4$)$_2$[TcCl$_6$] | O$_h$ | 10Dq = 4.5 |
| TcO$_2$ x H$_2$O | D$_{4h}$ | 10Dq = 4.5, Ds = -1.0 |
| (NBu$_4$)[TcOCl$_4$] | D$_{4h}$ | 10Dq = 4.25, Ds = -1.10, Dt = 0.28 |
| (NBu$_4$)[TcNCl$_4$] | D$_{4h}$ | 10Dq = 4.25, Ds = -1.10, Dt = 0.22 |
| KTcO$_4$ | T$_d$ | 10Dq = -2.5 |

**Tab. 1** The approximated local symmetry for Tc atoms and crystal-field parameters used in our calculations right hand side.

The Slater integrals obtained by Cowan's program, based on the Hartree–Fock method with relativistic corrections, were reduced to 80% of their Hartree-Fock values in present calculations. Note that an even larger reduction of the Slater integral values has been suggested[31] for 4d transition element solids versus molecular systems in these type of calculations to take into account the effect of the Tc 4d hybridization with ligand states. When we tried a larger reduction, it improved the agreement between calculated and experimental Tc $L_3$ XANES spectra for some compounds but made it worse for others. For consistent comparison between different the Tc oxidation states, we maintained a constant reduction of the Slater integral values to 80% throughout the whole series.

The approximated local symmetry for Tc atoms and crystal-field parameters used in the calculations for various compounds are summarized in Table 1. The calculated spectra were convoluted with Lorentzian (1.6 eV width) and Gaussian (0.5 eV width) functions to account for the core-hole broadening and instrumental resolution, respectively.[32] We refer to the ESI for a broader discussion concerning the theoretical calculations. Fig. 1 shows the Tc $L_3$ edge XANES spectra of all measured compounds, normalized to the maximum of the Tc $L_3$ absorption white line. The spectral shape of the white line arises from transitions from the $2p_{3/2}$ to 4d level. A systematic shift of the absolute white line position to higher values with increasing oxidation state is typically observed for many elements and at several edge transitions. This can be explained by the fact that, after oxidation, the core electrons become more tightly bound due to the reduced screening of the nuclear charge. The Tc ground state electron configuration is $[Kr]4d^55s^2$. When Tc ionizes, it loses its valence 5s electrons before losing 4d electrons. Tc has seven valence electrons that can be lost, but the number of 4d electrons will be different only for Tc with oxidation states higher than II. Therefore, the expected trend in the $L_3$ edge shift might be visible only for compounds with oxidation states from III to VII. The spectral shape is influenced by both the 4d occupancy as well as by the local symmetry and bond covalency of the ligands around the absorbing atom. These considerations are in agreement with our measurements. In Fig. 1 and Fig. 2 we show that the Tc $L_3$ white line shifts to higher energy with an increase of the oxidation state from III to VII (as much as one eV). The trend in shifts of the Tc $L_3$ white line is different for Tc systems with oxidation states lower than III, which might relate to both their similar number of 4d electrons in the ground state as well as to the variation in local symmetry for the Tc atoms. As the traditional methods of comparing the maximum of the white line or using the first edge inflection point seems to be inadequate, (Fig. S6), we used a center of gravity method as a more performant indicator for the Tc $L_3$ white line position (Fig. 2). The region of interest was chosen to start at 10 eV before the $TcO_2$ white line position and end at 5 eV after, as to include the main white line features. Altering the range by several eV's only effects the absolute values obtained slightly and preserves the trends observed. This way, with a minor degree of uncertainty, we report a correlation between formal oxidation state of the Tc atom and the edge shift. In Fig. 3 we show a representative set of experimental Tc $L_3$ edge XANES spectra along with their simulations by means of crystal-field multiplet theory. The center of gravity method applied to the simulations can be seen in Fig. 2. The calculations were performed for a series of Tc compounds with various oxidation states from II to VII, i.e. with different number of electrons in the 4d shell. The simulated spectra reproduce the main edge features well, demonstrating the strong influence of crystal-field interactions in the Tc 4d shell. $Tc^I$ has the same number of 4d electrons as $Tc^{II}$, and the presence of an extra-s-electron does not significantly influence the results of the calculations for the Tc $L_3$ spectra when comparing $4d^5$ and $4d^55s^1$ configurations. For those compounds, where either Cl or Br ligations were available, calculations were performed with Cl. The analogs with Br show narrower Tc $L_3$ spectra due to the – as expected - weaker crystal-field interaction and consequently reduced crystal-field splitting in accord with the spectrochemical series. Furthermore, for analogs with Br, the Slater integrals are expected to be further reduced as compared to those for compounds with Cl due to a stronger hybridization of Tc states with Br states.

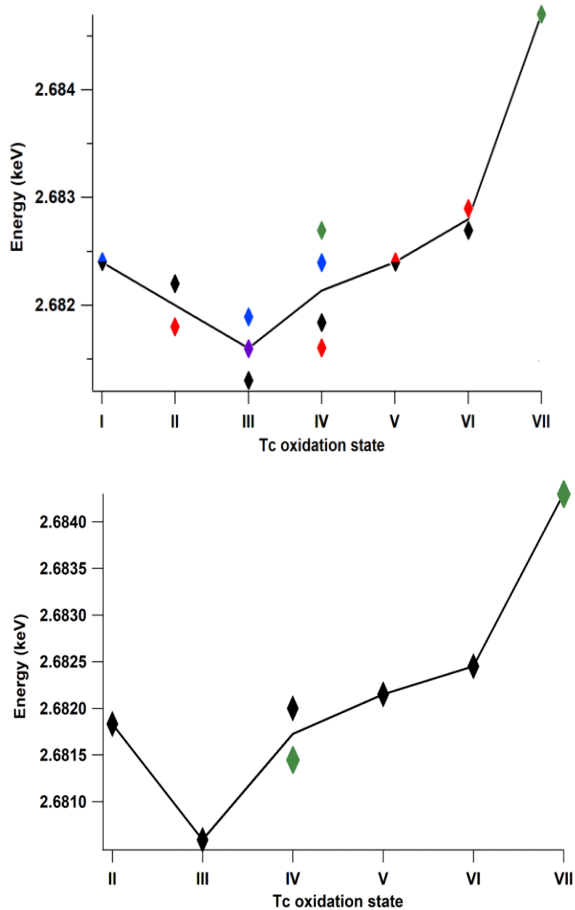

**Fig. 1** The trend of the white line position, as determined with the center of gravity method, for both the experimental (top) and calculated (bottom) data

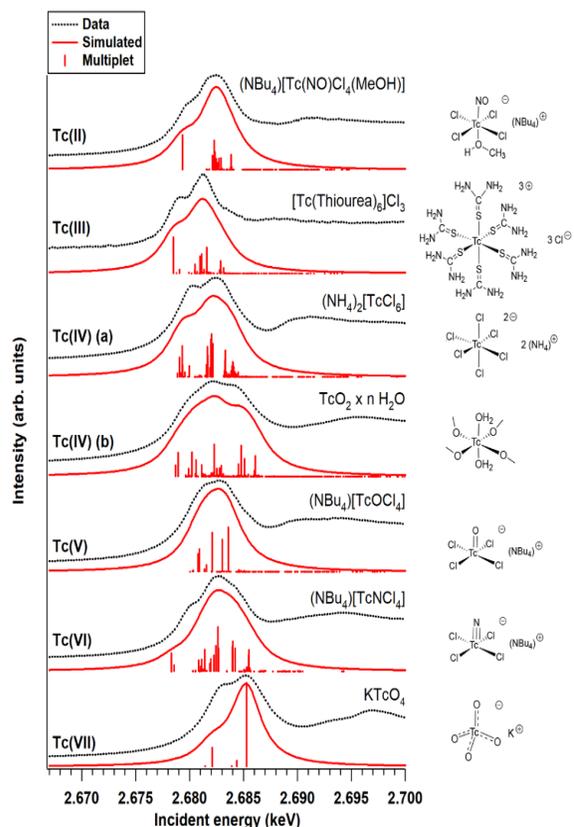

**Fig. 3** Representative increasing oxidation state series of Tc $L_3$ edge XANES and the simulated spectra based on crystal-field multiplet theory. A representation of the local geometry is provided for each compound on the right hand side.

Tc $L_3$ spectra of compounds with $PPh_3$ ligands reveal an additional feature at around 2685 eV. This feature can be attributed to the π-backbonding between Tc d and $PPh_3$.

The calculations show that the Tc $L_3$ spectral shape is affected by variations in the strong crystal-field interaction due to changes in local symmetry around Tc in different compounds. Therefore, a gravity center of the Tc $L_3$ lines needs to be used to consistently judge the chemical shifts for various oxidation states of Tc.

Recorded Tc $L_3$ spectra indeed reveal the expected chemical shifts towards the high-energy side when going from III to VII. On the other hand, no clear chemical shifts or slight variations are observed when going from I to III (Fig. 2). This could be due to two reasons: (i) the number of 4d electrons is the same in $Tc^I$ and $Tc^{II}$, only the 5s electron is removed; or (ii) the general chemical shift dependence on the change in the oxidation state may be affected by a switch to the low-spin ground state. The latter is well illustrated by calculations in Ref.[33] when a transition to the low-spin ground state for some $d^n$ configurations can lead to a significant change in the shape of the $L_3$ edge spectra.

As expected, the Tc $L_3$ edge is more sensitive to the oxidation state than Tc K edge XANES (probe of 5p electron level, which is empty for all Tc systems) because the $L_3$ white line position depends directly on the number of 4d electrons. Fig. S6 shows that there is virtually no shift of the Tc K edge white line from I to VII. Additionally, Tc $L_3$ XANES is found to be more sensitive to the Tc speciation in various compounds. It can be explained by the fact that 4d orbitals have a rather large radius and thus overlap strongly with orbitals of neighboring atoms. New experiments should follow similar strategies to record data on few reference systems and use electronic structure calculations to confirm the results.

In summary, the Tc oxidation state was measured by XANES at the Tc $L_3$ edge, for the first time on pure, unaltered Tc complexes. This method is currently the only one that can be used for Tc oxidation state characterization. We compared theoretical and experimental Tc $L_3$ edge XANES spectra for a comprehensive series of Tc compounds with varying oxidation states from I to VII. Crystal-field multiplet theory is well suited to simulate the Tc $L_3$ XANES structures and to give insights into the relevant parameters influencing the white line shape and position. Taking into account the overall conclusions arising from detailed analysis of the electronic structure of Tc compounds, we expect that such an approach can be extended to numerous studies of Tc systems and can stimulate further Tc coordination chemistry research relevant to various fields of applications.

## Conflicts of interest

There are no conflicts to declare.

## Acknowledgements

K.O.K and S.B acknowledge support from the European Research Council (ERC) (grant agreement No 759696). S. M. B. acknowledges support from the Swedish Research Council (research grant 2017-06465). N.M. acknowledges VESPA II project (02E11607B), supported by the German Federal Ministry of Economic Affairs and Energy (BMWi).

[a.] The Rossendorf Beamline at ESRF – The European Synchrotron, CS40220, 38043 Grenoble Cedex 9, France

[b.] Helmholtz-Zentrum Dresden-Rossendorf (HZDR), Institute of Resource Ecology, Bautzner Landstr. 400, 01328 Dresden, Germany

[c.] Karlsruhe Institute of Technology (KIT), Institute for Nuclear Waste Disposal (INE), P.O. Box 3640, D-76021 Karlsruhe, Germany

[d.] Freie Universität Berlin, Institute of Chemistry and Biochemistry, Fabeckstr. 34/36, D-14195 Berlin, Germany

[e.] Molecular and Condensed Matter Physics, Department of Physics and Astronomy, Uppsala University, P.O. Box 516, Uppsala, Sweden